%% file: main.tex
\documentclass[journal]{IEEEtran}
\usepackage{amsmath,amsfonts}

\usepackage[algo2e,ruled,vlined,linesnumbered]{algorithm2e}
\SetInd{0em}{0.5em}
\SetKwComment{Comment}{/\!\!/}{}

\usepackage{array}
\usepackage{textcomp}
\usepackage{stfloats}
\usepackage{url}
\usepackage{verbatim}
\usepackage{graphicx}
\usepackage{cite}
\usepackage{subcaption}
\usepackage{hyperref}
\hypersetup{%
    colorlinks = false,
    hidelinks
    }

\usepackage{pgf}
\usepackage{pgfplots}
\usepackage{tikz}
\usetikzlibrary{spy}
\pgfplotsset{compat=1.18}

\usepackage{svg}
\usepackage{epstopdf}
\usepackage{physics}
\usepackage{makecell}

\usepackage{array}
\newcolumntype{H}{>{\setbox0=\hbox\bgroup}c<{\egroup}@{}}
\usepackage[normalem]{ulem}

\usepackage{format} %
\newcommand{\ObsOp}{\bE}

\definecolor{burgundy}{rgb}{0.5, 0.0, 0.13}

\begin{document}
\title{Learned Primal Dual Splitting for Self-Supervised Noise-Adaptive MRI Reconstruction}

\author{%
Nikola Janju\v{s}evi\'{c}$^1$, 
Amirhossein Khalilian-Gourtani$^{2}$, 
Yao Wang$^3$,
Li Feng$^1$ 
\thanks{%
New York University Grossman School of Medicine, %
\{%
$^1$Radiology Department, %
$^2$Neurology Department,%
\},
New York, NY 10016, USA.%
}
\thanks{%
$^3$New York University Tandon School of Engineering, %
Electrical and Computer Engineering Department, %
Brooklyn, NY 11201, USA.%
}
\thanks{Please send all correspondence regarding to this manuscript to N. Janju\v{s}evi\'{c} (email:nikola.janjusevic@nyulangone.org).}
\vspace*{-20pt}}%

\markboth{\LaTeX~Template 2024}%
{Shell \MakeLowercase{\textit{et al.}}: A Sample Article Using IEEEtran.cls for IEEE Journals}

\maketitle

\begin{abstract}
Magnetic resonance imaging (MRI) reconstruction has largely been dominated by deep neural networks (DNN); however, many state-of-the-art architectures use black-box structures, which hinder interpretability and improvement. Here, we propose an interpretable DNN architecture for self-supervised MRI reconstruction and denoising by directly parameterizing and learning the classical primal-dual splitting, dubbed LPDSNet. This splitting algorithm allows us to decouple the observation model from the signal prior. Experimentally, we show other interpretable architectures without this decoupling property exhibit failure in the self-supervised learning regime. We report state- of-the-art self-supervised joint MRI reconstruction and denoising performance and novel noise-level generalization capabilities, where in contrast black-box networks fail to generalize.
\end{abstract}


\bstctlcite{IEEEexample:BSTcontrol}

\thispagestyle{empty} %
\input{sections/intro.tex}

\thispagestyle{empty}
\input{sections/proposed.tex}

\thispagestyle{empty}
\input{sections/results.tex}

\thispagestyle{empty}
\section{Discussion and Conclusion}
We presented LPDSNet, an interpretable architecture for noise-level robust self-supervised MRI reconstruction. LPDSNet improves upon 
other interpretable architectures~\cite{janjusevicCDLNet2022} by decoupling the observation model from transform-domain sparsity prior, leading to substantially improved supervised learning performance, and crucially enabling self-supervised learning where CDLNet~\cite{janjusevicCDLNet2022} fails. Though \soa networks also make use of observation-prior decoupling, use of black-box constructions render them parameter-inefficient and unable to generalize to unseen noise-levels. These qualities make LPDSNet an attractive candidate for MRI reconstruction on limited-sample noisy datasets without clean images, and for further adaptation to other imaging inverse problems.

\section*{Acknowledgments}
This work was supported in part by National Institute of Biomedical Imaging and Bioengineering (NIBIB) grants: R01EB030549, R01 EB031083 and R21 EB032917. Authors thank NYU Langone Health UltraViolet for technical support. 

\ifCLASSOPTIONcaptionsoff%
  \newpage
\fi
\bibliographystyle{IEEEtran}
\bibliography{IEEEabrv,references}

\end{document}

%% file: sections/intro.tex
\section{Introduction and Background}\label{sec:intro}

Magnetic resonance imaging (MRI) is a powerful clinical tool for in-vivo
imaging, offering the advantage of non-ionizing radiation, which eliminates
associated side effects and provides a safe way to view the patient's body.
However, MRI scanners are often slow due to the inherent limitations of
magnetization physics. Hence, MR images are typically reconstructed from
under-sampled data (in Fourier-space)~\cite{fastmri}. To enable the development of more
affordable scanners and increase their accessibility, there is growing demand
for MRI use in low signal-to-noise ratio (SNR) regimes where signal recovery
presents additional challenges.

Deep Neural Networks (DNNs) are the \soa for MRI reconstruction. However, these \soa DNN
architectures are most often constructed as black-boxes, rendering analysis and
improvement beholden to trial-and-error. 
So-called ``unrolled networks'' attempt to bring interpretability to DNN
architectures by taking a classical optimization algorithm, truncating it to
$K$-iterations, and parameterizing each iteration with learnable functions.
However, most unrolled networks parameterize their iterations/layers with
existing black-box architectures, again hindering analysis and improvement \cite{janjusevicCDLNet2022, janjusevicGDLNet2022}.
Previous network architectures based on primal-dual splitting have been
proposed in the literature. However, these networks either employ black-box sub-networks~\cite{Adler2018, rudzusika2021invertible},
introduce complicated weight parameterizations~\cite{LPDnetZhang2021}, and/or employ approximations
which deviate from the classical algorithm~\cite{LPDnetZhang2021}. 

In this work, we introduce a DNN architecture for 
noisy MRI reconstruction that is a direct parameterization of the primal-dual splitting algorithm with 
an $\ell_1$ sparsity penalty in an analysis-transform domain. To this end, the
proposed network is initialized \textit{exactly} as the classical algorithm,
without approximation. 
We show our proposed network, LPDSNet, to be \soa in
joint MRI reconstruction and denoising, with novel noise-level generalization
capabilities in the self-supervised MRI reconstruction task.

\subsection{Classical MRI Reconstruction}
We model our observations as a subsampled multicoil Fourier domain signal ${\y \in \C^{N_sC}}$, with $C$ coils, of a ground-truth image ${\x \in \C^N}$,
with fourier-sampling locations denoted by index-set $\Omega$ and coil-sensitivity maps ${\bs \in \C^{NC}}$, 
\begin{equation}
\y_c = \IDMAT_{\Omega}\bF( \bs_c \circ \x ) + \bnu_c, \quad \forall~c=1,\dots,C,
\end{equation}
where $\bF$ is the $N$-pixel 2D-DFT matrix, $\circ$ denotes element-wise multiplication, and
${\IDMAT_{\Omega} \in \{0,1\}^{N_s\times N}}$ is the row-removed identity matrix
with kept rows indicated by $\Omega$. For simplicity, we consider our
observations to be noise-whitened such that ${\bnu_c \sim \N(0,
\sigma^2\IDMAT)}$ is per-channel i.i.d. Gaussian noise with noise-level
$\sigma$ identical in each channel. We denote this observation model through
the encoding operator $\ObsOp$ defined such that ${\y = \ObsOp\x + \bnu}$. We
say the observation is accelerated by a factor of ${N/N_s}$.

A classical signal-processing framework for MRI reconstruction is regularization 
via $\ell_1$-sparsity in an analysis-transform domain, given as,
\begin{equation} \label{eq:analysis-sc}
\min_{\x} \frac{1}{2}\norm{\y - \ObsOp\x}_2^2 + \lambda \norm{\bD^H\x}_1,
\end{equation}
where $\bD^H$ is some given linear analysis-transform.
An effective method for solving~\eqref{eq:analysis-sc} is via 
primal-dual splitting,
\begin{equation} \label{eq:primaldual}
\min_{\x}\max_{\z} \frac{1}{2}\norm{\y - \ObsOp\x}_2^2 + \RE\{\z^H\bD^H\x\}  - \I_{\lambda \B_1}(\z),
\end{equation}
where $\I_{\lambda\B_1}$ is the indicator function of the $\lambda$-$\ell_1$-norm ball.
We can solve~\eqref{eq:primaldual} via gradient descent w.r.t.\ the primal variable ($\x$)
and proximal gradient ascent w.r.t.\ the dual-variable ($\z$)~\cite{chambolle_ctsopt}, 
known as the Primal-Dual Splitting method (PDS),\footnote{a.k.a.\ the Chambolle-Pock algorithm, the Primal Dual Hybrid Gradient method. This specific variant is often referred to as the Condat-Vu PDS method \cite{chambolle_ctsopt}.}
\begin{equation} \label{eq:pds}
\begin{aligned}
    \x^{(k+1)} &= \x^{(k)} - \eta\left( \ObsOp^H \left( \ObsOp\x - \y \right) + \bD\z \right), \\
    \tilde{\x}^{(k+1)} &= \x^{(k+1)} + \theta(\x^{(k+1)} - \x^{(k)}), \\
    \z^{(k+1)} &= \Pi_{\lambda}\left( \z^{(k)} + \beta \bD^H \tilde{\x}^{(k+1)} \right),
\end{aligned}
\end{equation}
where the above iterates, $k$, are repeated until convergence. Here, $\theta$ is an
extrapolation parameter in $[0,1]$ and step-sizes $\eta$ and $\beta$ must
satisfy ${(\tfrac{1}{\eta}-L_E)\tfrac{1}{\beta} \geq \norm{\bD}^2_2}$ for convergence, 
where ${L_E = \norm{\ObsOp}_2 \approx 1}$ for common parallel MRI observation models.
The dual-update uses 
an element-wise magnitude clipping function for complex input vector,
${\Pi_\lambda(\z) = \sign(\z)\min(\abs{\z},~\lambda)}$, where $\sign(z=re^{j\phi}) = e^{j\phi}$.
In general, one may consider learning an optimal 
transform $\bD^H$ over a dataset to achieve better reconstruction.

\subsection{The Convolutional Dictionary Learning Network (CDLNet)} 
Originally formulated for natural image restoration, and recently extended to supervised MRI reconstruction in \cite{janjusevic2024groupcdl}, CDLNet~\cite{janjusevicCDLNet2022}
is a directly parameterized unrolled network which derives its architecture from a synthesis-transform domain sparsity based proximal-gradient descent. The CDLNet architecture for MRI reconstruction is given as~\cite{janjusevic2024groupcdl},
\begin{gather} \label{eq:cdlnet}
    \tilde{\y} = \ObsOp^H\y - \mu,\quad\quad\z^{(0)} = 0, \nonumber \\
    \text{\small for $k=0:K-1$} \hskip\textwidth \nonumber \\
    \begin{array}{|lcrcl} 
        \z^{(k+1)} &=& \ST_{\btau^{(k)}}\left( \z^{(k)} - {\bA^{(k)}}^H\ObsOp^H\left(\ObsOp{\bB^{(k)}}\z^{(k)} - \tilde{\y} \right) \right),
    \end{array} \nonumber \\
    \hat{\x} = f_\Theta(\y;\,\ObsOp) = \bD\z^{(K)} + \mu, 
\end{gather}
where ${\ST_{\tau}(\z) = \sign(\z)\max(0, \abs{\z} - \tau)}$ is the element-wise soft-thresholding operator and ${\mu = \mathrm{mean}(\ObsOp^H\y)}$. In total, the 
parameters of CDLNet are given by ${\Theta = (\bD, \, \{ {\bA^H}^{(k)}, \, {\bB}^{(k)}, \, \btau^{(k)} \}_{k=0}^{K-1})}$, 
where $\bA,\bB,\bD$ are learned convolutional operators.
This architecture was shown to perform on par with \soa black-box networks for
supervised MRI reconstruction in~\cite{janjusevic2024groupcdl}. 
CDLNet's synthesis-transform sparsity prior
couples the learned convolutions with the MRI encoding operator, $\bE$. 
Experimentally, we show this makes CDLNet sensitive to observation model shifts between training
and inference (see Section \ref{sec:results:jrd}) and is the source of
catastrophic failure for CDLNet in self-supervised MRI reconstruction techniques
such as SSDU~\cite{yaman2020self_mrm} (described in Section
\ref{sec:intro:ssdu}).

\subsection{The End2End Variational Network (E2EVarNet)}
E2EVarNet~\cite{sriram2020end} is a \soa MRI reconstruction network 
which derives itself from unrolling $K$ iterations of gradient descent. Consider the recovery of a ground-truth fourier-space sample, $\bk$, as ${\min_{\bk} \tfrac{1}{2}\sum_{c=1}^C\norm{\bM\left(\bk_c - \IDMAT^T_\Omega\y_c\right)}_2^2 + g(\bk)}$ with regularizer $g$. The E2EVarnet architecture unrolls gradient-descent,
\begin{gather} \label{eq:varnet}
    \bk^{(k+1)} = \bk^{(k)} - \eta\left( \bM(\bk^{(k)} - \bk^{(0)}) + \nabla g(\bk^{{k}}) \right),
\end{gather}
where a UNet~\cite{unet} (image-domain black-box DNN) is used to implement
the gradient of the regularizer, $\nabla g$, with separate learned weights for each
iteration. 
Here, ${\bk^{(0)} = \IDMAT_\Omega^T\y}$ is the zero-filled observed
fourier-space sample, and ${\bM = \IDMAT_\Omega^T\IDMAT_\Omega}$ is the
fourier-space sampling mask for fourier-space signal $\y$ sampled with index set
$\Omega$. 
This use of black-box networks potentially leads to a parameter-inefficient design structure that is unable to tackle issues with noise-level mismatch between training and inference (see Section \ref{sec:results:noisegen}).

\subsection{DNN Training and Inference with SSDU}\label{sec:intro:ssdu}
Consider a dataset of noisy subsampled fourier-space observations and corresponding encoding operators, 
${\D = \{ (\y_\Omega,~\ObsOp_\Omega)_{i} \}_{i=1}^{N_d}}$, where set $\Omega$ denotes the fourier sampling indices for each observation.
The Self-Supervision by Data-Undersampling (SSDU) framework~\cite{yaman2020self_mrm} trains a DNN for MRI reconstruction 
by splitting given fourier-space samples into an input set $\Xi$ and a target
set $\Lambda$, such that $\Omega = \Xi \cup \Lambda$.
With ${(\y_\Omega,~\ObsOp_\Omega)}$ as our original fourier-space observation and encoding operator, we denote input and target 
samples as ${(\y_\Xi,~\ObsOp_\Xi)}$ and ${(\y_\Lambda,~\ObsOp_\Lambda)}$, respectively. We then train our neural network with parameters $\Theta$, $f_\Theta$, using the following loss,
\begin{equation}
    \min_\Theta \sum_{(\y_\Omega,\,\ObsOp_\Omega)\in \D} \Loss\left( \y_\Lambda,\,\ObsOp_\Lambda f_\Theta(\y_\Xi; \ObsOp_\Xi) \right),
\end{equation}
where $\Loss$ is any differentiable loss function and we assume $f_\Theta$ produces a complex-valued image-domain estimate.
At inference, we input the entire fourier-space sample to obtain our image estimate,
${\hat{\x} = f_\Theta(\y_\Omega; \ObsOp_\Omega)}$. 
This discrepancy between the encoding operator used by the network during
training ($\ObsOp_\Xi$) and inference ($\ObsOp_\Omega$) necessitates the network to be stable w.r.t.\  
changes in the observation model. As we show in Section~\ref{sec:results:jrd}, CDLNet~\cite{janjusevicCDLNet2022} fails when presented with this observation model mismatch.

%% file: sections/proposed.tex
\section{Proposed Method}
\subsection{The Learned Primal Dual Splitting Network (LPDSNet)}
We propose unrolling PDS~\eqref{eq:pds} for $K$ iterations 
as follows,
\begin{gather} \label{eq:lpdsnet}
    \tilde{\y} = \ObsOp^H\y - \mu,\quad\x^{(0)} = 0,\quad\z^{(0)} = 0, \nonumber\\
    \text{\small for $k=0:K-1$} \hskip\textwidth \nonumber \\
    \begin{array}{|rcl} 
        \x^{(k+1)} &=& \x^{(k)} - \eta^{(k)} \left( \ObsOp^H\ObsOp\x^{(k)} - \tilde{\y} + \bB^{(k)}\z^{(k)} \right) \\
        \tilde{\x}^{(k+1)} &=& \x^{(k+1)} + \theta^{(k)}(\x^{(k+1)} - \x^{(k)}) \\
        \z^{(k+1)} &=& \Pi_{\blambda^{(k)}} \left(\z^{(k)} + {\bA^H}^{(k)}\tilde{\x}^{(k+1)} \right) 
    \end{array} \nonumber \\
    \hat{\x} = f_\Theta(\y;\, \ObsOp) = \x^{(K)} + \mu,
\end{gather}
where $\mu=\mathrm{mean}(\ObsOp^H\y)$. We parameterize the analysis-operator
($\bD^H$) and its adjoint ($\bD$) as ($1 \rightarrow M$)-channel and ($M
\rightarrow 1$)-channel $s$-strided convolution analysis and synthesis operators (${\bA^H}^{(k)}$, ${\bB}^{(k)}$), respectively,
with learnable complex valued kernels of size $p \times p$ for each layer. 
We learn step-size parameters $\eta^{(k)} \in \R_+$ and 
subband-dependent clipping parameters $\blambda^{(k)} \in \R_+^M$, for all layers.

Altogether the parameters of the network are given by, 
${\Theta = \{ {\bA^H}^{(k)}, \, {\bB}^{(k)}, \, \eta^{(k)}, \, \theta^{(k)}, \, \blambda^{(k)} \}_{k=0}^{K-1}}$. 
Inspired by the noise-adaptive thresholding of CDLNet, we employ
noise-adaptive clipping in LPDSNet by parameterizing the learned clipping
values as a function of the estimated input noise-level ($\hat{\sigma}$), i.e. 
${\blambda^{(k)} = \blambda_0^{(k)} + \blambda_1^{(k)}\hat{\sigma}}$. This noise-level may be estimated via filtering methods (as in \cite{janjusevicCDLNet2022}) or via a reference scan.

The proposed architecture contains the hallmarks of
successful black-box DNN architectures, such as convolutions, point-wise nonlinearities,
and residual connections. However, LPDSNet notably lacks
feature-domain-to-feature-domain learned operators\footnote{All conv ops in
LPDSNet go to and from the image-domain and a latent subband domain.}, does
not use deep-learning tricks such as batch-norm or layer-norm, and uses a
clipping nonlinearity ($\Pi_\lambda$) uncharacteristic of typical DNNs.
Though the magnitude-clipping
nonlinearity is not differentiable everywhere for real-valued inputs, for
complex-valued inputs it is differentiable almost everywhere. As such,
LPDSNet directly parameterizes unrolled primal-dual splitting without
the use of black-box deep-learning elements and without the use of
approximations for differentiability.

%% file: sections/results.tex
\section{Results} \label{sec:results}
\noindent{\bf Dataset}:
We use the publicly available FastMRI T2-weighted Brain Dataset \cite{fastmri},
which we contaminate with synthetic additive Gaussian white noise (AWGN). We
generate a training dataset by using a single mask and noise-level $(\sigma \in
\sigmatrain)$ per volume. The masks are generated with random cartesian
subsampling of fourier-space rows at $4\times$ acceleration with $8\%$ of center
lines kept. We use ESPIRiT~\cite{espirit} to generate ground-truth
coil sensitivity maps from the noise-free multicoil samples. We limit our
training dataset to volumes with 20 coils, yielding 83 training volumes and 28
test volumes.

\noindent{\bf Training}:
We train each network (CDLNet, LPDSNet, E2EVarNet) for a total of 300k
gradient-steps using the Adam optimizer~\cite{adam} with batch-size of 1, a cosine-annealing learning
rate starting at $5\times10^{-4}$ and ending at $2\times 10^{-6}$, and with otherwise
default parameters. We project the parameters of CDLNet and LPDSNet after each
gradient-step to ensure non-negative step-sizes, thresholds, and clip-values. As in \cite{yaman2020self_mrm}, we use a
normalized mixed $\ell_1$-$\ell_2$ loss in the image-domain and fourier-domain
for supervised and self-supervised learning, respectively. We use the SSDU loss~\cite{yaman2020self_mrm}
for self-supervised learning with a uniform 2D subsampling rate of $80\%$ and a $10\times10$ region around the center of fourier-space preserved.

\noindent{\bf Architecture}:
We use the same hyperparamters for CDLNet and LPDSNet architectures: ${K=20}$
layers, ${M=64}$ subbands, stride-$2$ convolutions with size $7\times 7$ kernels, and noise-adaptive thresholding/clipping, respectively. For E2EVarNet, we use the
ground-truth sensitivity maps and augment the network to always output
complex-valued image-domain results using said sensitivity maps. We
follow the hyperparameters given in~\cite{sriram2020end} and implementation
provided by~\cite{fastmri}, though using only ${K=6}$ layers instead of ${K=12}$ as
in the original implementation. We emphasize that the results of each of the
presented methods may be improved by increasing network size, except in cases of catastrophic failure and lack of noise-level generalization.

\noindent{\bf Initialization of LPDSNet}:
We initialize analysis and synthesis convolutional dictionaries with the same
filters, $\bD$, which have been spectrally normalized, i.e.
(${\bA^{(k)}=\bB^{(k)}=\bD~\forall~k}$) and ${\norm{\bD}_2 = 1}$.
Following~\cite{chambolle_ctsopt}~(Alg. 8), we initialize step-size
${\eta^{(k)}=1/2}$, noting that typical MRI observation operators have spectral norms close to unity. We initialize learned clipping values as $\blambda_0=10^{-3}$ and $\blambda_1=0$.

\noindent{\bf Implementation}:
We implemented CDLNet, LPDSNet, and E2EVarNet in the Julia Programming Language using Lux.jl~\cite{pal2023lux}.
The E2EVarNet implementation follows closely the publicly available code in~\cite{fastmri}. We trained each model on a single NVIDIA A100 GPU for
approximately 48 hours.

\subsection{Joint Reconstruction and Denoising}\label{sec:results:jrd}

\begin{figure}[tbh]
    \centering
    \includegraphics[width=0.5\textwidth]{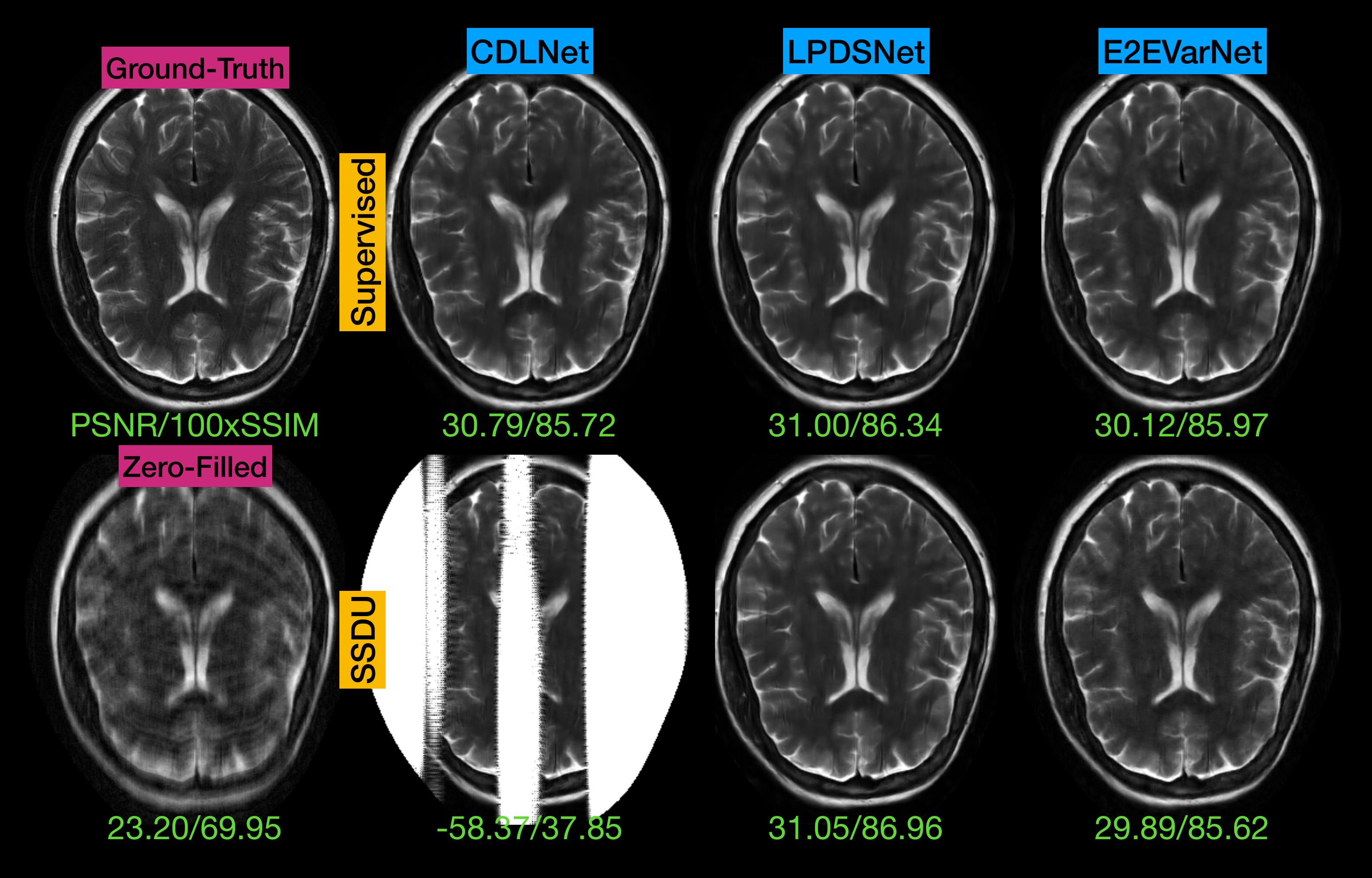}
    \caption{%
        Visual comparison of models trained for ${4\times}$ acceleration with
        ${\sigmatrain = [0.04,0.06]}$, evaluated on a test-set slice with
        ${\sigmatest = 0.05}$. Magnitude images, clipped to max-value of 1, shown
        for display purposes. All inferences used the same fourier-space mask
        and noise realization. Note that CDLNet's catastrophic failure doesn't happen for every sample, but happens for a large proportion, such as shown above.
    }
    \label{fig:compare05}
\end{figure}

We first evaluate the performance of LPDSNet in the joint MRI reconstruction and denoising task. When trained with a supervised loss, LPDSNet achieves nearly a 2 dB increase in PSNR compared to the widely used E2EVarNet\cite{sriram2020end}, while using orders of magnitude fewer parameters (Tbl.\ref{tbl:PSNR_SSIM_JRD}, Fig.\ref{fig:compare05}). When using self-supervised training, CDLNet catastrophically fails to reconstruct images during inference (much lower PSNR compared to zero-filled, Tbl.\ref{tbl:PSNR_SSIM_JRD} and Fig.\ref{fig:compare05}). This failure might be caused by the mismatch in mask sampling properties between training and inference, leading to numerical instability for CDLNet at inference time. In contrast, both LPDSNet and E2EVarNet remain stable in the SSDU scenario. Notably, LPDSNet trained with a self-supervised loss achieves the best performance (Tbl.\ref{tbl:PSNR_SSIM_JRD}, Fig.\ref{fig:compare05}), slightly outperforming the supervised case. Note that self-supervision can enhance robustness against overfitting when limited noisy samples are available during training.

\begin{table}[htb]
\centering
\caption{%
    PSNR/$100\times$SSIM for networks trained under supervised and self-supervised
    joint denoising and MRI reconstruction tasks on the FastMRI T2w Brain Dataset~\cite{fastmri} with ${\sigmatrain = [0.04,0.06]}$,
    ${\sigmatest=0.05}$, and ${4\times}$ acceleration. Learned parameter counts shown below method names. $(^\ast)$ highlights catastrophic failure compared to zero-filled {PSNR/$100\times$SSIM=$26.85/79.23$}.
}
\resizebox{\linewidth}{!}{%
\begin{tabular}{ccHcc} \hline
Loss & 
\makecell{CDLNet~\cite{janjusevicCDLNet2022} \\ 262k} & 
\makecell{AltCDLNet~\\ 288k} & 
\makecell{LPDSNet~\\ 240k} &
\makecell{E2EVarNet~\cite{sriram2020end}\\ 15M} 
\\ \hline
    Supervised & 29.82/87.66 & 32.05/88.49 & 32.53/89.09 & 30.80/88.40 \\ 
    SSDU       & 12.87$^{(\ast)}$/81.89 & 32.00/88.89 & \textcolor{black}{\bf 32.58}/\textcolor{black}{\bf 89.51} & 30.75/88.16 \\
\hline
\end{tabular}%
}
\label{tbl:PSNR_SSIM_JRD}
\end{table}

\subsection{Noise-level Generalization} \label{sec:results:noisegen}
\begin{figure}[tbh]
    \centering
    \includegraphics[width=0.5\textwidth]{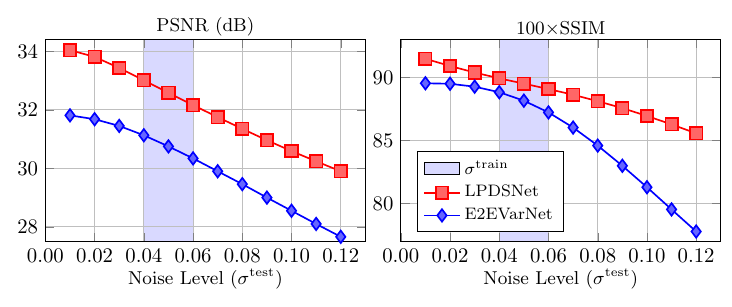}
    \caption{%
        Noise-level generalization of SSDU-trained networks for joint MRI reconstruction and denoising of $4\times$ accelerated FastMRI T2w Brain Data~\cite{fastmri}.
    }
    \label{fig:noisegen}
\end{figure}

\begin{figure}[tbh]
    \centering
    \includegraphics[width=0.5\textwidth]{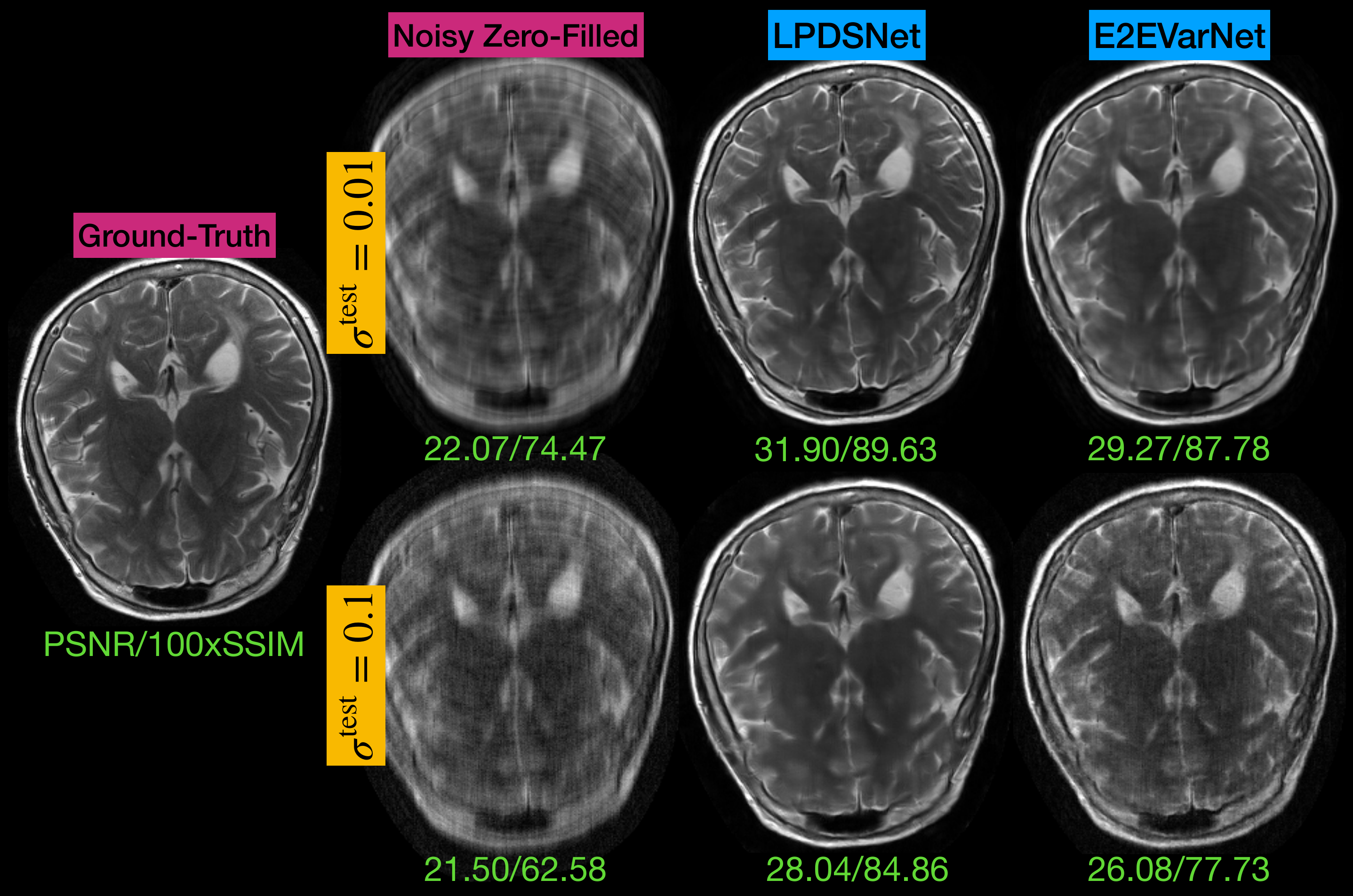}
    \caption{%
        Visualization of noise-level generalization of the proposed LPDSNet. 
        A single LPDSNet and E2EVarNet~\cite{sriram2020end} were trained for ${4\times}$
        acceleration with ${\sigmatrain \in [0.04,0.06]}$ using the SSDU loss. 
        Magnitude images shown for display purposes. The same mask
        and noise realization are used between all inferences, with the noise
        realization scaled appropriately to ${\sigmatest = 0.01}$ and ${\sigmatest
        = 0.1}$ in the top and bottom rows, respectively.
    }
    \label{fig:noisegen_vis}
\end{figure}

Next we investigate the robustness of different architectures when there is a mismatch between training and inference noise-levels. CDLNet first showed such property in the context of image denoising, however, in the context of joint MRI reconstruction and denoising such an experiment is not meaningful as CDLNet shows failure at inference in the self-supervised scenario. We trained both E2EVarNet and our proposed LPDSNet over the noise-range ${\sigmatrain \in [0.04,0.06]}$ using the SSDU loss. We then tested these networks over a wider noise-level $\sigmatest\in[0.01,0.12]$. We observe superior performance of LPDSNet in all noise-levels. In stark contrast to E2EVarNet, our proposed architecture generalizes to noise-levels not present during training (note the plateau in E2EVarNet PSNR/SSIM in $\sigmatest\in[0.01,0.04]$ and sharp decrease in SSIM for $\sigmatest\in[0.06,0.12]$ in Fig.\ref{fig:noisegen} as well as Fig.\ref{fig:noisegen_vis} for visual comparison). We highlight that LPDSNet decouples the data consistency term from the sparsity prior, resulting in numerical stability during self-supervised training as well as robustness to unseen noise-levels during training given the noise-adaptive clipping used in the architecture.
\vspace{-1em}
